# Superhalogens as Building Blocks of Complex Hydrides for Hydrogen Storage

Ambrish Kumar Srivastava, Neeraj Misra*

Department of Physics, University of Lucknow, Lucknow, 226007, India

*Corresponding author, E-mail: neerajmisra11@gmail.com

**Abstract**

Superhalogens are species whose electron affinity (EA) or vertical detachment energy (VDE) exceed to those of halogen. These species typically consist of a central electropositive atom with electronegative ligands. The EA or VDE of species can be further increased by using superhalogen as ligands, which are termed as hyperhalogen. Having established $BH_4^-$ as a superhalogen, we have studied $BH_{4-x}(BH_4)_x^-$ ($x$ = 1−4) hyperhalogen anions and their Li-complexes, $LiBH_{4-x}(BH_4)_x$ using density functional theory. The VDE of these anions is larger than that of $BH_4^-$, which increases with the increase in the number of peripheral $BH_4$ moieties ($x$). The hydrogen storage capacity of $LiBH_{4-x}(BH_4)_x$ complexes is higher but binding energy is smaller than that of $LiBH_4$, a typical complex hydride. The linear correlation between dehydrogenation energy of $LiBH_{4-x}(BH_4)_x$ complexes and VDE of $BH_{4-x}(BH_4)_x^-$ anions is established. These complexes are found to be thermodynamically stable against dissociation into $LiBH_4$ and borane. This study not only demonstrates the role of superhalogen in designing new materials for hydrogen storage, but also motivates experimentalists to synthesize $LiBH_{4-x}(BH_4)_x$ ($x$ = 1−4) complexes.

## 1. Introduction

Hydrogen energy has been recognised as a clean, efficient and eco-friendly alternative to fossil fuels [1]. Complex metal hydrides are considered as promising materials for hydrogen storage [2], which are composed of metal cation and a complex anion. Typical examples include alanates [$M(AlH_4)_x$] and borohydrides [$M(BH_4)_x$], where M is a metal atom with valence *x*. Borohydrides such as $LiBH_4$, $NaBH_4$, $Ca(BH_4)_2$, $Mg(BH_4)_2$ and $Al(BH_4)_3$ etc. have been well recognized for their hydrogen storage behaviour. Due to high gravimetric and volumetric densities, $LiBH_4$ is a good choice for efficient hydrogen storage [2]. $LiBH_4$ is a salt due to charge transfer from $Li^+$ to $BH_4^-$. The $BH_4^-$ complex anion is stabilized by the distribution of extra electron over H atoms, which is reflected in its high vertical detachment energy (VDE) or electron affinity (EA) exceeding to that of halogen atom. Such species whose EAs or VDEs exceed to that of Cl (3.80 eV) [3] are termed as superhalogens [4, 5]. Superhalogens are hypervalent species, which need an extra electron to complete their octet and stabilize them. These species have been continuously reported in literature theoretically as well as experimentally [6-14]. The use of superhalogen for hydrogen storage has been proposed only recently [15].

According to Gutsev and Boldyrev [4], general formula for a typical superhalogen anion is $MX_{k+1}^-$, where ligand X is an electronegative atom and *k* is the nominal valence of central electropositive atom M. Since the valence of B is 3 and therefore, $BH_4^-$ is a superhalogen. The EA or VDE of a typical superhalogen can be further increased by employing superhalogen as ligands instead of electronegative atom. For instance $AlF_4^-$ is a superhalogen with VDE of 9.79 eV, which can be enhanced to 12.05 eV for $Al_5F_{16}^-$ in which all F atoms are replaced with $AlF_4$ moieties [16]. Such superhalogens are termed as hyperhalogens [17]. In the same way, one can expect the design of new superhalogen from $BH_4^-$ in which H atoms are successively substituted with $BH_4^-$ moieties. In this communication, we present the design of $BH_{4-x}(BH_4)_x^-$ ($x$ = 1−4) hyperhalogens using density functional theory. We have discussed the hydrogen storage behaviour of their $LiBH_{4-x}(BH_4)_x$ complexes and derived an interesting relation between the dehydrogenation energy of $LiBH_{4-x}(BH_4)_x$ and VDE of their anions.

## 2. Computational details

All lithium complexes [$LiBH_{4-x}(BH_4)_x$] and corresponding anions [$BH_{4-x}(BH_4)_x^-$] considered in this study were fully optimized at ωB97xD method [18] using 6-311+G(d) basis set in Gaussian 09 program [19]. The ωB97xD method incorporates long range dispersion correction [20], which has recently been used for hydrogen storage on small clusters [21]. The geometry optimization was performed without any symmetry constraints and followed by frequency calculations to ensure that the optimized structures correspond to true minima in the potential energy surface. The vertical detachment energy (VDE) of anions has been calculated by difference of total energy of optimized structure of anions and corresponding neutral structure:

$$\mathrm{VDE} = E[BH_{4-x}(BH_4)_x]_{\mathrm{single\ point}} - E[BH_{4-x}(BH_4)_x^-]_{\mathrm{optimized}} \quad (x = 0-4) \quad \ldots\ldots..(1)$$

where $E[BH_{4-x}(BH_4)_x^-]_{\mathrm{optimized}}$ is the total energy of optimized structure of $BH_{4-x}(BH_4)_x^-$ and $E[BH_{4-x}(BH_4)_x]_{\mathrm{single\ point}}$ is the single point energy of neutral structure at optimized geometry of corresponding anion. The binding energy ($E_b$) and dehydrogenation energy ($\Delta E$) of $LiBH_{4-x}(BH_4)_x$ complexes are calculated by considering their dissociation $LiBH_{4-x}(BH_4)_x \rightarrow Li^+ + BH_{4-x}(BH_4)_x^-$ and $LiBH_{4-x}(BH_4)_x \rightarrow LiH + m\,B + n\,H_2$, respectively.

$$E_b = E[Li^+] + E[BH_{4-x}(BH_4)_x^-] - E[LiBH_{4-x}(BH_4)_x] \quad \ldots\ldots\ldots(2)$$

$$\Delta E = E[LiH] + m\,E[B] + n\,E[H_2] - E[LiBH_{4-x}(BH_4)_x] \quad (x = 0-4) \quad \ldots\ldots\ldots(3)$$

where $E$[..] represent total electronic energy of respective species including zero point correction. In order to explore the stability of $LiBH_{4-x}(BH_4)_x$ against dissociation into $LiBH_4$

+ $x$ $BH_3$, we have calculated corresponding dissociation energy ($D_e$) and enthalpy ($\Delta H$) using following equations:

$D_e = E[LiBH_4] + x\,E[BH_3] - E[LiBH_{4-x}(BH_4)_x]$

$\Delta H = \Delta H[LiBH_4] + x\,\Delta H[BH_3] - \Delta H[LiBH_{4-x}(BH_4)_x]$ ($x$ = 1−4) ……….(4)

where $\Delta H$[..] is thermal enthalpy of respective species calculated at 298.15 K.

## 3. Results and discussion

Any attempt to develop complex hydrides as practical hydrogen storage materials requires understanding of their electronic structure and the energetics of their fundamental dehydrogenation and rehydrogenation processes. We start our discussion considering $LiBH_4$, whose crystal structure was first determined by Harris and Meibohm [22]. The crystal structure along with optimized structure of $BH_4^-$ and $LiBH_4$ are displayed in Fig. 1. In crystal form, each $Li^+$ ion is surrounded by four $BH_4^-$ ions in a tetrahedral configuration such that $BH_4^-$ is strongly deformed. This feature is also reflected in the optimized structure of $BH_4^-$ and $LiBH_4$. Note that $BH_4^-$ assumes a tetrahedral structure with bond length of 1.243 Å. In $LiBH_4$, however, two B−H bond lengths vary significantly from 1.210 Å to 1.264 Å. The VDE of $BH_4^-$ is calculated to be 4.50 eV as listed in Table 1. In order to design $BH_{4-x}(BH_4)_x^-$ hyperhalogen ($x$ = 1−4), we have replaced H atoms in $BH_4^-$ superhalogen by $BH_4$ moieties successively. The optimized structures of $BH_{4-x}(BH_4)_x^-$ hyperhalogens are shown in Fig. 2. The VDEs of these hyperhalogens are larger than that of $BH_4^-$ superhalogen, which increase with the increase in peripheral $BH_4$ moieties and reach to as high as 7.28 eV for $B_5H_{16}^-$ in which all H atoms are replaced with $BH_4$. Owing to large EA of $BH_4$ as compared to H atom, the extra electron is easily delocalized over peripheral $BH_4$. This delocalization increases with the increase in $BH_4$ moieties, which results in increase in the VDE of corresponding anions. We have also displayed the equilibrium structures of Li-complexes of these hyperhalogens in Fig. 2. One can see that Li atom interacts with four, three, two and one H atoms of $BH_{4-x}(BH_4)_x$ hyperhalogen for $x$ = 1, 2, 3 and 4, respectively. Table 1 lists the binding energy ($E_b$) of $Li^+$ and $BH_{4-x}(BH_4)_x^-$ ions in $LiBH_{4-x}(BH_4)_x$ calculated by equation (2). One can note that the $E_b$ of $LiBH_{4-x}(BH_4)_x$ decreases with the increase in the VDE of $BH_{4-x}(BH_4)_x^-$ hyperhalogen anions. This can also be expected due to delocalization of extra electron over several $BH_4$ moieties.

Table 1 also lists hydrogen capacity (in % weight) of $LiBH_{4-x}(BH_4)_x$ complexes. Note that the hydrogen content of $LiBH_4$ 18.51 wt% is higher than that of $LiAlH_4$ (10.62 wt%), which makes it preferable for hydrogen storage. The hydrogen content of $LiBH_{4-x}(BH_4)_x$ complexes is larger than that of $LiBH_4$, which increases monotonically with the increase in $x$, i.e., from 19.81 wt% ($x$ = 1) to 20.91 wt% ($x$ = 4). Therefore, the proposed $LiBH_{4-x}(BH_4)_x$ complexes should be preferable over $LiBH_4$ due to high hydrogen storage capacity. The overall dehydrogenation of $LiBH_4$ proceeds as $LiBH_4 \rightarrow LiH + B + 3/2\ H_2$ such that $LiBH_4$ liberates three of the four hydrogen atoms upon melting at 280 °C and decomposes into LiH and boron [23]. Analogous to $LiBH_4$, we have considered the dehydrogenation of $LiBH_{4-x}(BH_4)_x$ into LiH + $m$ B + $n$ $H_2$ and calculated the dehydrogenation energy ($\Delta E$) using equation (3). The calculated $\Delta E$ values of $LiBH_{4-x}(BH_4)_x$ complexes including $LiBH_4$ are also listed in Table 1. For $LiBH_{4-x}(BH_4)_x$, this value increases from 14.58 eV for $x$ = 1 to 31.92 eV for $x$ = 4. As mentioned earlier, $LiBH_{4-x}(BH_4)_x$ complexes are stabilized due to charge transfer from Li to $BH_{4-x}(BH_4)_x$ hyperhalogens. The thermodynamic stability of a series of metal borohydrides, $M(BH_4)_n$ (where M is a metal with valence $n$) has been studied by first principle calculations [24]. It has been found that have the heat of formation of $M(BH_4)_n$ depends linearly on the Pauling electronegativity of M. In case of $LiBH_{4-x}(BH_4)_x$ complexes, we notice that the $\Delta E$ varies linearly with the VDE of $BH_{4-x}(BH_4)_x^-$. The correlation plot of $\Delta E$ and VDE is available upon request. The correlation between $\Delta E$ and VDE follows a linear equation with the correlation coefficient, $R^2$ = 0.98

$\Delta E = 8.81\,(\mathrm{VDE}) - 32.85$ ………..(5)

Although $LiBH_{4-x}(BH_4)_x$ complexes possess high capacity of hydrogen storage, it is important to analyze their kinetic and thermodynamic stability against dissociation into $LiBH_4$ and borane ($BH_3$). Table 1 lists the dissociation energy ($D_e$) and dissociation enthalpy ($\Delta H$) of $LiBH_{4-x}(BH_4)_x$ complexes calculated by equations (4). One can note that both $D_e$ and $\Delta H$ are positive for $x = 1-4$ and therefore, all complexes are kinetically and thermodynamically stable. Furthermore, their stability increases with the increase in the VDE of hyperhalogen anions. This may suggest that $LiBH_{4-x}(BH_4)_x$ can be synthesized using $LiBH_4$ and borane at least via gas phase reaction. Recently, the potential of complex hydrides for electrochemical energy storage has been experimentally demonstrated [25]. Likewise, the complex hydrides reported here may be employed for electrochemical storage as well.

## 4. Conclusions

We have exploited the concept of superhalogen to design hyperhalogens anions, $BH_{4-x}(BH_4)_x^-$ ($x = 1-4$) and their Li-complexes, $LiBH_{4-x}(BH_4)_x$. The interaction between Li and $BH_{4-x}(BH_4)_x$ is similar to that between Li and $BH_4$, i.e., ionic. The binding energy of complexes decreases but VDE of these anions increases with the increase in $x$, the number of peripheral $BH_4$ moieties due to delocalization of extra electron over several $BH_4$ moieties. The hydrogen storage capacity of $LiBH_{4-x}(BH_4)_x$ complexes is higher than that of $LiBH_4$, a typical complex hydride. The dehydrogenation energy of $LiBH_{4-x}(BH_4)_x$ complexes is found to be linearly related to the VDE of $BH_{4-x}(BH_4)_x^-$. We have also demonstrated the thermodynamic stability of these complexes against dissociation into $LiBH_4$ and borane.

## Acknowledgement

A. K. Srivastava acknowledges Council of Scientific and Industrial Research, India for a research fellowship [Grant No. 09/107(0359)/2012-EMR-I].

**References**

[1] A. M. Seayad, D. M. Antonelli, Adv Mater. 16 (2004) 765–777.

[2] S. Orimo, Y. Nakamori, J. R. Eliseo, A. Zuttel, Craig M. Jensen, Chem. Rev. 107 (2007) 4111-4132.

[3] H. Hotop, W. C. Lineberger, J. Phys. Chem. Ref. Data 14 (1985) 731.

[4] G. L. Gutsev, A. I. Boldyrev, Chem. Phys. 56 (1981) 277-283.

[5] G. L. Gutsev, A. I. Boldyrev, Chem. Phys. Lett. 108 (1984) 250-254.

[6] G. L. Gutsev, J. Chem. Phys. 98 (1993) 444-452.

[7] X. -B. Wang, C. -F. Ding, L. -S. Wang, A. I. Boldyrev, J. Simons, J. Chem. Phys. 110 (1999) 4763−4771.

[8] A. N. Alexandrova, A. I. Boldyrev, Y. -J. Fu, X. Yang, X. -B. Wang, L. -S. Wang, J. Chem. Phys. 121 (2004) 5709 –5719.

[9] B. M. Elliott, E. Koyle, A. I. Boldyrev, X.-B. Wang, L.-S. Wang, J. Phys. Chem. A 109 (2005) 11560 – 11567.

[10] J. Yang, X. -B. Wang, X. -P. Xing, L. -S.Wang, J. Chem. Phys. 128 (2008) 201102.

[11] G. L. Gutsev, C. A. Weatherford, K. Pradhan, P. Jena, J. Comput. Chem. 32 (2011) 2974.

[12] M. Marchaj, S. Freza, P. Skurski, Chem. Phys. Lett. 560 (2013) 15-21.

[13] A. K. Srivastava, N. Misra, Chem. Phys. Lett. 624 (2014) 15–18.

[14] A. K. Srivastava, N. Misra, New J. Chem. 39 (2015) 9543–9549.

[15] P. Jena, J. Phys. Chem. Lett. 6 (2015) 1119-1125.

[16] C. Sikorska, P. Skurski, Chem. Phys. Lett. 536 (2012) 34-38.

[17] M. Willis, M. Gotz, A. K. Kandalam, G. F. Gantefor, P. Jena, Angew. Chem. Int. Ed. 49 (2010) 8966-8970.

[18] J. -D. Chai, M. Head-Gordon, Phys. Chem. Chem. Phys. 10 (2008) 6615 – 6620.

[19] M. J. Frisch, G. W. Trucks, H. B. Schlegel, et al. Gaussian 09, Revision B.01, Gaussian Inc., Wallingford, CT (2010).

[20] S. Grimme, J. Comput. Chem. 27 (2006) 1787−1799.

[21] R. Shinde, M. Tayade, J. Phys. Chem. C 118 (2014) 17200-17204.

[22] P. M. Harris, E. P. Meibohm, J. Am. Chem. Soc. 69 (1947) 1231.

[23] D. S. Stasinevich, G. A. Egorenko, Russ. J. Inorg. Chem. 13 (1968) 341.

[24] Y. Nakamori, K. Miwa, A. Ninomiya, H. W. Li, N. Ohba, S. Towata, A. Zuttel, S. Orimo, Phys. Rev. B 74 (2006) 45126.

[25] A. Unemoto, M. Matsuo, S. Orimo, Adv. Funct. Mater. 24 (2014) 2267–2279.

Table 1. The VDE of $BH_{4-x}(BH_4)_x^-$ anions, binding energy ($E_b$), hydrogen content (H), dehydrogenation energy ($\Delta E$), dissociation energy ($D_e$) and enthalpy ($\Delta H$) of $LiBH_{4-x}(BH_4)_x$ complexes obtained at  at ωB97xD/6-311+G(d) level.

| $BH_{4-x}(BH_4)_x^-$ | | $LiBH_{4-x}(BH_4)_x$ | | | | | | |
|---|---|---|---|---|---|---|---|---|
| $x$ | VDE (eV) | $E_b$ (eV) | H (%wt) | Dehydrogenation[a] | | | Dissociation[b] | |
| | | | | $m$ | $n$ | $\Delta E$ (eV) | $D_e$ (eV) | $\Delta H$ (eV) |
| 0 | 4.50 | 6.62 | 18.51 | 1 | 3/2 | 7.85 | - | - |
| 1 | 5.52 | 6.34 | 19.81 | 2 | 3 | 14.58 | 1.50 | 1.34 |
| 2 | 6.25 | 6.15 | 20.38 | 3 | 9/2 | 20.78 | 2.46 | 2.16 |
| 3 | 6.64 | 5.80 | 20.71 | 4 | 6 | 26.48 | 2.93 | 2.54 |
| 4 | 7.28 | 5.45 | 20.91 | 5 | 15/2 | 31.92 | 3.14 | 2.65 |

[a]dehydrogenation path, $LiBH_{4-x}(BH_4)_x \rightarrow LiH + m\,B + n\,H_2$

[b]dissociation path,  $LiBH_{4-x}(BH_4)_x \rightarrow LiBH_4 + x\,BH_3$

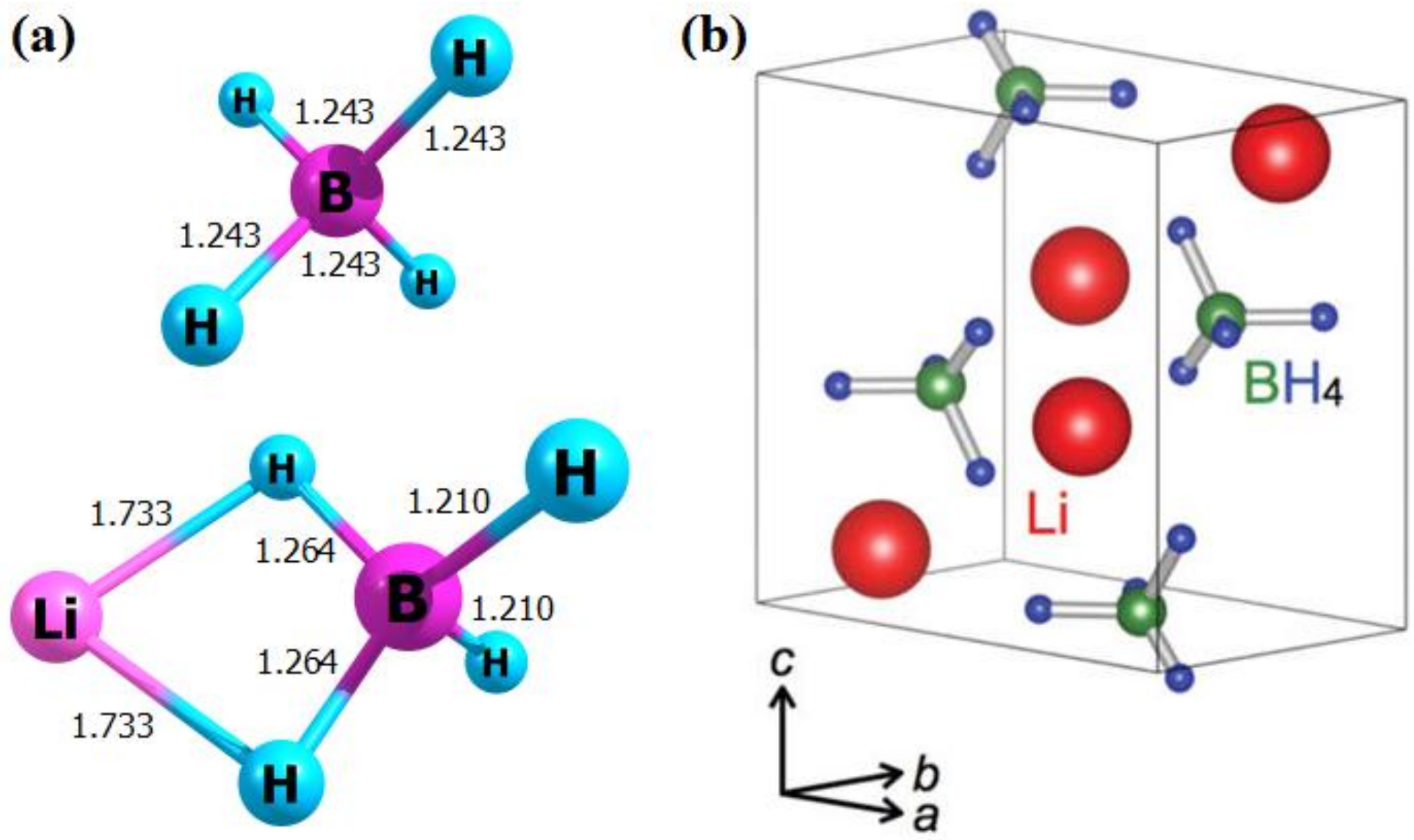

Fig. 1. (a) Equilibrium structures of $BH_4^-$ and $LiBH_4$ obtained at ωB97xD/6-311+G(d) level with bond lengths (in Å), and (b) crystal structure of $LiBH_4$ taken from ref. [15].

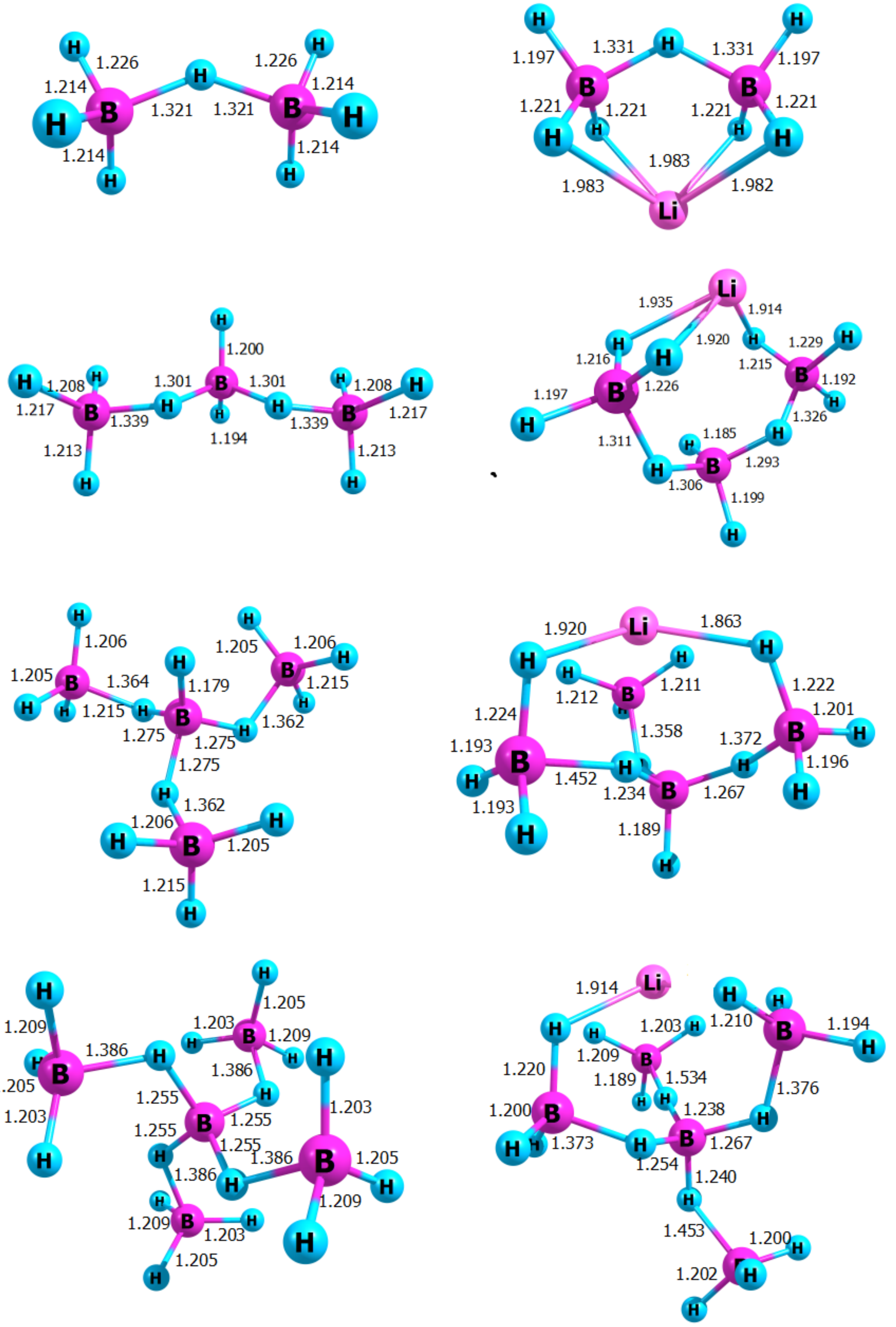

Fig. 2. Equilibrium structures of $BH_{4-x}(BH_4)_x^-$ hyperhalogen anions (left) and $LiBH_{4-x}(BH_4)_x$ complexes (right) obtained at ωB97xD/6-311+G(d) level for $x$ = 1−4. Bond lengths (in Å) are also given.